\documentclass[prb,twocolumn,amsmath,amssymb,floatfix,superscriptaddress,aps]{revtex4-1}
\usepackage[utf8]{inputenc}
\usepackage{graphicx}
\usepackage{dcolumn}
\usepackage{bm}
\usepackage{xcolor}
\usepackage{soul}
\usepackage{cancel}

\newcommand{\be}{\begin{equation}}
\newcommand{\ee}{\end{equation}}
\newcommand{\bea}{\begin{eqnarray}}
\newcommand{\eea}{\end{eqnarray}}
\newcommand{\ba}{\begin{align}}
\newcommand{\ea}{\end{align}}

\def\a{\alpha}

\def\g{\gamma}
\def\G{\Gamma}

\def\D{\Delta}

\def\w{\omega}

\def\bra{\langle}
\def\ket{\rangle}

\def\Tr{{\rm Tr}}

\def\1op{\hat{\mathbbm{1}}}

\setstcolor{blue}

\renewcommand{\vr}{{\bf r}}

\begin{document}

\title{Steady-state density functional theory for thermoelectric effects}

\author{N. Sobrino}
 \email{nahualcsc@dipc.org}
\affiliation{Donostia International Physics Center, Paseo Manuel de Lardizabal 4, E-20018 San Sebasti\'an, Spain}
\affiliation{Nano-Bio Spectroscopy Group and European Theoretical Spectroscopy Facility (ETSF), Departamento de F\'isica de Materiales, Universidad del Pa\'is Vasco UPV/EHU, Avenida de Tolosa 72, E-20018 San Sebasti\'an, Spain}

 \author{R. D'Agosta}
\email{roberto.dagosta@ehu.es}
 \affiliation{Nano-Bio Spectroscopy Group and European Theoretical Spectroscopy Facility (ETSF), Departamento de F\'isica de Materiales, Universidad del Pa\'is Vasco UPV/EHU, Avenida de Tolosa 72, E-20018 San Sebasti\'an, Spain}
\affiliation{IKERBASQUE, Basque Foundation for Science, Maria Diaz de Haro 3, E-48013 Bilbao, Spain}

\author{S. Kurth}
\email{stefan.kurth@ehu.es}
\affiliation{Nano-Bio Spectroscopy Group and European Theoretical Spectroscopy Facility (ETSF), Departamento de F\'isica de Materiales, Universidad del Pa\'is Vasco UPV/EHU, Avenida de Tolosa 72, E-20018 San Sebasti\'an, Spain}
\affiliation{IKERBASQUE, Basque Foundation for Science, Maria Diaz de Haro 3, E-48013 Bilbao, Spain}
\affiliation{Donostia International Physics Center, Paseo Manuel de Lardizabal 4, E-20018 San Sebasti\'an, Spain}

\date{\today}
\begin{abstract}
The recently proposed density functional theory for steady-state
transport (i-DFT) is extended to include temperature gradients between
the leads. Within this framework, a general and exact expression is
derived for the linear Seebeck coefficient which can be written as the sum of
the Kohn-Sham coefficient and an exchange-correlation contribution.
The formalism is applied to the single-impurity Anderson model for
which approximate exchange-correlation functionals are suggested for
temperatures both above and below the Kondo temperature.  A certain
structural property of the exchange-correlation potentials in the Coulomb
blockade regime allows to recover an earlier result expressing the
Seebeck coefficient in terms of quantities of equilibrium density functional
theory.  The numerical i-DFT results are compared to calculations
with the numerical renormalization group over a wide range of
temperatures finding a reasonable agreement while i-DFT comes at a
much lower computational cost.
\end{abstract}

\maketitle

\section{introduction}
During the last years, thermoelectric materials have attracted the attention
of both experimental and theoretical research interest due to their wide
variety of possible technological
applications.\cite{Vining:NAT:2008,Vining:NAT:2009,Dubi:RMP:2011}
These applications aim mainly at the conversion of waste heat into
an electrical current through the Seebeck effect or at cooling through
the Peltier effect.\cite{goldsmid:2010:introduction} An efficient thermocouple
for heat-to-electricity conversion exhibits a low thermal conductance
as well as a large electrical conductance and Seebeck 
coefficient.\cite{goldsmid:2010:introduction} In standard
bulk materials there are intrinsic difficulties in disentangling the phonon
and electron contributions to the transport coefficients: improving the
electrical conductance usually increases the thermal conductance and
reduces the Seebeck coefficient. Systems of reduced dimensionality are a
promising alternative since a more accurate and independent control of the
diverse factors dominating the thermoelectric efficiency appears
possible.\cite{Hicks1993} The path
towards molecular electronics and the discovery of two dimensional
materials further expand the scope of potential materials for thermoelectric
applications.\cite{Datta1997,Cuniberti2005,CuevasScheer:10} However, it still
remains a difficult problem to predict which new materials (bulk, layered or
heterostructure) are most apt as thermoelectric devices.

The theory of thermoelectric energy conversion draws from many
sources,\cite{Beenakker:91,Zianni2008,Sothmann2013,Sanchez2013a,DAgosta2013,Eich2014a}
but recently, Density Functional Theory (DFT) has acquired a prominent
role in almost any material
modeling.\cite{Cuniberti2005,Hohenberg1964,Kohn1965}
This popularity is mainly due to two factors: (i) relatively
simple approximations to the so-called exchange-correlation (xc) functional,
the central quantity of DFT, lead to reasonably accurate results for many
material properties and (ii) DFT is numerically very efficient. 
For the description of (electronic or thermal) transport, DFT is typically
combined with the Landauer-B\"uttiker (LB-DFT)
formalism.\cite{Datta1997,Cuniberti2005,CuevasScheer:10,ThossEvers:18}
However, formally this approach is incomplete since it treats electrical
transport as a scattering problem of non-interacting electrons and uses
equilibrium DFT for an inherently out-of-equilibrium system. A proper
DFT framework to describe electronic transport is time-dependent DFT (TDDFT).
\cite{RungeGross:84} In TDDFT, one may view steady-state transport as the
long-time limit of a system initially in equilibrium and driven out of
equilibrium by application of an external bias. In the steady state, this
approach will then lead to equations structurally identical to the LB-DFT
equations but with the external bias augmented by a dynamical xc correction.
\cite{StefanucciAlmbladh:04,StefanucciAlmbladh:04-2,SaiZwolakVignaleDiVentra:05,KoentoppBurkeEvers:06,StefanucciKurthRubioGross:06,VignaleDiVentra:09}
As a consequence, the connection between quantities of the real system and those
calculated within the LB-DFT framework is typically nontrivial and strong
corrections may appear.\cite{Stefanucci:NL:2015,Yang:PRB:2016}

More recently, an alternative DFT framework to describe steady-state
electronic transport, dubbed i-DFT, has been
proposed.\cite{Stefanucci:NL:2015} i-DFT directly focusses on the steady state
and is designed to give both the density and the current of a (DC) biased
molecular junction.\cite{Stefanucci:NL:2015} However, it does not make any
statements on the time evolution towards such a steady state. Nevertheless,
just as TDDFT, it also leads to xc corrections to the bias. 
So far, functionals for these xc bias corrections have been constructed
for model systems.\cite{Stefanucci:NL:2015,Kurth:PRB:2016,JacobKurth:18}
In particular, i-DFT allows to describe the single impurity Anderson model
(SIAM) both in the Coulomb Blockade (CB) as well as in the Kondo regime.
\cite{Kurth:PRB:2016}

The standard form of i-DFT assumes that the
device and the leads are thermally equilibrated. In this work we
generalize the formalism to finite temperature gradients
between the leads which allows one to use i-DFT as a highly efficient method
to calculate the Seebeck coefficient. Paying special attention to the
temperature dependence of the i-DFT equations we derive: a) an exact expression
for the linear Seebeck coefficient for any system and any regime in terms of pure i-DFT quantities and b) a pair of
xc functionals for the SIAM with explicit dependence on the
thermal gradient $\Delta T$ between the leads which are exact for
$T\gg \gamma$, i.e., for weakly coupled leads. We also model the
effect of the broadening due to the couplings in the functionals and
we equip our functionals to describe the linear Seebeck coefficient in
the Kondo regime in an accurate way.

\section{i-DFT for Thermoelectric effects}

We consider the typical transport setup where a central region, e.g.,
a single molecule or a quantum dot, is coupled to a left (L) and a
right (R) electrode. The system is driven out of equilibrium by 
applying a DC bias $V$ across the junction and we are interested in
the resulting steady-state current $I$. The recently suggested i-DFT 
framework for steady-state transport\cite{Stefanucci:NL:2015} is based on a
one-to-one mapping between, on the one hand, the density $n(\vr)$ in
the central region and the steady current $I$ through it and, on the
other hand, the external potential $v(\vr)$ in the same region and the
bias $V$ across it. In the original formulation, both left and right
leads are kept at the same temperature $T$ which enters the formalism
as an external parameter only. Here we propose an extension of i-DFT
to include a temperature difference between the leads. This thermal
gradient creates an electronic current which can be compensated by a
bias in an open circuit setup and thus allows to study the Seebeck
effect. For simplicity, we symmetrically apply both a bias $V$ as well
as a temperature difference $\D T$ between the two leads, i.e., we
have $V_{\a}=\pm V/2$ and $T_{\a}=T\pm\D T/2$ where $\a=L,R$.
\footnote{Needless to say, we assume $T>\Delta T/2$ such that $T_{L},T_{R}>0$.} Of
course, now both $T_L$ and $T_R$ (or, equivalently, $T$ and $\D T$)
enter as parameters into the formalism. If we make the (physically
reasonable) assumption that the density in the central region and the
current are continuously differentiable at $\D T=0$, the original
i-DFT proof \cite{Stefanucci:NL:2015} of the one-to-one correspondence
between ``densities'' and ``potentials'' can directly be applied to
our situation and we can formulate the i-DFT theorem for leads at
different temperatures (see also
Ref.~\onlinecite{KurthJacobSobrinoStefanucci:19}).

{\em Theorem}: 
For any pair of finite temperatures $T_{\alpha}$ in the leads, there
exists a one-to-one correspondence between the pair of ``densities''
$(n,I)$ and the pair of ``potentials'' $(v,V)$ in a finite (and gate
dependent) region around zero voltage $V$ and zero thermal gradient
$\Delta T$.

It is important to note that the proof of the theorem goes through for any
form of the interaction, particularly also for the non-interacting case.
As usual, in order to establish a Kohn-Sham (KS) scheme, we have to assume
non-interacting representability, i.e., that the same densities $(n,I)$ of an
interacting system (with potentials $(v,V)$) can also be obtained as
densities of a non-interacting system with potentials $(v_s,V_s)$. 
Following the standard KS procedure, we define the
Hartree-exchange-correlation (Hxc) gate potential as
$v_{\rm Hxc}[n,I]=v_s[n,I]-v[n,I]$  and the xc bias as 
$V_{\rm xc}[n,I]=V_s[n,I]-V[n,I]$. The self-consistent coupled KS equations for the
density and the current then are
\ba
n(\vr)&=2\sum_{\alpha=L,R}\int \frac{{\rm d}\w}{2 \pi} f_{\alpha}(\w-V_{\alpha,s})
A_{\alpha,s}(\vr,\w)
\label{n_KS}\\
I&=2\sum_{\alpha=L,R}\int \frac{{\rm d}\w}{2 \pi}
f_{\alpha}(\w-V_{\alpha,s})s_{\alpha}\mathcal{T}(\w)
\label{I_KS}
\end{align}
where $f_{\alpha}(x)=1/(e^{x/T_{\alpha}}+1)$ is the Fermi function (for
lead $\a$), $s_{L/R}=\pm$, and $V_{\alpha,s}=s_{\a}(V+V_{\rm xc})/2$.
$A_{\alpha,s}(\vr,\w)=\bra\vr\rvert\mathcal{G}(\w)\Gamma_{\alpha}(\w)
\mathcal{G}^{\dagger}(\w)\rvert\vr\ket$ is the partial KS spectral function
with the KS Green's function $\mathcal{G}$ and the broadening matrix
$\Gamma_{\alpha}$  of lead $\alpha$. Finally, the transmission function is
given by $\mathcal{T}(\w)=\Tr\left[\mathcal{G}(\w)\Gamma_{L}(\w)
  \mathcal{G}^{\dagger}(\w)\Gamma_{R}(\w)\right]$.

Eqs.~(\ref{n_KS}) and (\ref{I_KS}) have the same structure as the original i-DFT
equations \cite{Stefanucci:NL:2015} with the exception that the temperature
difference $\Delta T$ between the two leads enters explicitly both in the Fermi
functions $f_{\a}$ and in the functionals for $v_{\rm Hxc}$ and $V_{\rm xc}$. In
the following we use exactly this property to derive an exact expression for
the Seebeck coefficient in terms of i-DFT quantities.

The Seebeck coefficient is defined as that bias which has to be applied to
compensate a small temperature difference between the leads such that no
current flows. Formally it can be written as 
\be
S=\left.\frac{dV}{d\Delta T}\right\rvert_{\substack{V=0\\ \Delta T=0}}=
\left.\frac{dI/d\Delta T}{dI/dV}\right\rvert_{\substack{V=0\\ \Delta T=0}}.
\label{S_def}
\ee

Both the numerator and the denominator of Eq.~(\ref{S_def}) can be calculated
directly from Eq.~(\ref{I_KS}). The denominator is nothing but the zero-bias
conductance $G$ which can be expressed as \cite{Stefanucci:NL:2015}
\be
\label{dI_dV}
G=\left.\frac{dI}{dV}\right\rvert_{\substack{V=0\\ \Delta T=0}}=\frac{G_{s}}{1-G_{s}\left.\frac{\partial V_{xc}}{\partial I}\right\rvert_{{\substack{V=0\\ \Delta T=0}}}}
\ee
where we have defined the KS zero bias conductance
\be
G_{s}=-\int \frac{{\rm d}\omega}{2 \pi} f'(\omega)\mathcal{T}(\omega) \;.
\label{ks_cond}
\ee
where $f'(x) = {\rm d}f/{\rm d}x$ with the Fermi function
taken at the temperature $T=T_L=T_R$.

Similarly, the numerator of Eq.~(\ref{S_def}) can be calculated as 
\bea
\label{dI_dDT}
\left.\frac{dI}{d\Delta T}\right\rvert_{\substack{V=0\\ \Delta T=0}} &=&
\int \frac{{\rm d}\omega}{2 \pi} \left[f'(\omega)\left(\frac{\omega}{T}+\frac{dV_{xc}}{d\Delta T}\right)\right]\mathcal{T}(\omega)\nonumber \\
&=&\frac{L_{s}-G_{s}\left.\frac{\partial V_{xc}}{\partial \Delta T}\right\rvert_{{\substack{V=0\\ \Delta T=0}} }}{1-G_{s}\left.\frac{\partial V_{xc}}{\partial I}
  \right\rvert_{{\substack{V=0\\ \Delta T=0}}}}
\eea
where we have defined
\be
L_{s}=\frac{1}{T}\int \frac{{\rm d}\omega}{2 \pi} f'(\omega)\omega
\mathcal{T}(\omega) \;.
\ee
In deriving Eq.~(\ref{dI_dDT}) we have expanded out the total
derivative,
\be
\frac{{\rm d}V_{xc}}{d\Delta T}= \frac{\partial V_{xc}}{\partial \Delta T} 
+ \frac{\partial V_{xc}}{\partial I} \frac{dI}{d\Delta T} 
+\int {\rm d}^3 r\; \frac{\delta V_{xc}}{\delta n(\vr)} 
\frac{d n(\vr)}{d\Delta T},
\ee
and used the fact that for $I=0$ (i.e., $V=0$ and $\Delta T = 0$) the last term
vanishes because $V_{\rm xc}[n,I=0]=0$. 

Combining Eqs.~(\ref{dI_dV}) and (\ref{dI_dDT}), we then arrive at the
following simple expression for the linear Seebeck coefficient
\be
S=S_{s}-\left.\frac{\partial V_{xc}}{\partial \Delta T}
\right\rvert_{\substack{\Delta V=0\\ \Delta T=0}},
\label{linear_Seebeck_dft}
\ee
where $S_{s}=\frac{L_{s}}{G_{s}}$ is KS Seebeck coefficient and the second term
of Eq.~(\ref{linear_Seebeck_dft}) is the xc contribution. 
Eq.~(\ref{linear_Seebeck_dft}) is one of the central results of the present
work. It is formally exact and expresses the Seebeck coefficient of a general
interacting system solely in terms of i-DFT quantities. In practice, of course,
one has to use approximations for the Hxc gate and the xc bias functionals.
We will address the construction of such approximations for a model system in
the next Section. We also point out that an expression similar to
Eq.~(\ref{linear_Seebeck_dft}) was recently obtained in a standard DFT
framework\cite{Yang:PRB:2016} in the Coulomb Blockade (CB) regime. In the
following Section, we will discuss the connection between the two approaches.

\section{Single Impurity Anderson Model}
\label{siam}

In this Section we consider the single-impurity Anderson model (SIAM) which
also in previous works \cite{Stefanucci:NL:2015,Kurth:PRB:2016,Yang:PRB:2016}
has been used as a first model for the development of approximate i-DFT
functionals. The SIAM describes a single interacting impurity level (quantum
dot) coupled to a left (L) and right (R) lead. The dot is described by
the Hamiltonian
\be
\hat{H}^{\rm dot} = \sum_{\sigma} v \hat{n}_{\sigma} + U \hat{n}_{\uparrow}
\hat{n}_{\downarrow}
\ee
where $v$ is the on-site energy of the dot and $U$ is the interaction and
$\hat{n}_{\sigma}$ is the operator for the density of electrons with spin
$\sigma$ on the dot. The system is coupled to left and right featureless
electronic leads described by frequency-independent
couplings $\G_{\a}(\w)=\g_{\a}$ (with $\a=L,R$), i.e., we work in the
wide band limit (WBL). The leads are characterized by temperature $T_{\a}$ and
may be subject to a DC bias $V_{\alpha}$ which we take to be symmetric, i.e.,
$V_L=-V_R=V/2$. 

If we want to study the system in an i-DFT framework, we need approximations
for the (H)xc functionals. The approximate functionals designed in previous
work \cite{Stefanucci:NL:2015,Kurth:PRB:2016} were restricted to the case of
equal lead temperatures, $T_L=T_R$, and therefore we need to extend the
construction to the more general case $T_L\neq T_R$.

{\em Coulomb blockade regime - }Following ideas used in earlier work
\cite{Stefanucci:NL:2015}, we first aim to construct approximations for the
xc functionals in the Coulomb blockade regime. We start by 
expressing both the density on and the current through the dot
in terms of the many-body spectral function $A(\w)$:
\begin{subequations}
  \be
    \begin{split}
      n =& \int \frac{{\rm d}\w}{2 \pi}\left[ \frac{2\g_L}{\g}  f_L(\w-\mu-V_L)\right.\\
     &+ \left. \frac{2\g_R}{\g}  f_R(\w-\mu-V_R) \right] A(\w)
\end{split}
\ee
\be
\begin{split}
  I =& \frac{2 \g_L \g_R}{\g} \int \frac{{\rm d}\w}{2 \pi}\; \left[ f_L(\w-\mu-V_L) \right. \\
    & \left .- f_R(\w-\mu-V_R) \right]
A(\w)
\end{split}
\ee
\label{dens_curr_siam}
\end{subequations}
where $\g=\g_L+\g_R$ is the total broadening. 

We want to use Eqs.~(\ref{dens_curr_siam}) to reverse engineer the (H)xc
potentials of i-DFT, therefore we need a model for the many-body spectral
function $A(\w)$. As a starting point we use the exact spectral function of the
isolated dot which is given by
\be
A_0^{\rm mod}(\omega)=\left(1-\frac{n}{2}\right)\delta(\omega-v)+
\frac{n}{2}\delta(\omega-v-U) \;.
\label{A_delta_model}
\ee
Using $A_0^{\rm mod}(\omega)$ as model spectral function for the
{\em contacted} dot and inserting it into Eqs.~(\ref{dens_curr_siam}) leads
to exactly the same expressions for density and current as one would obtain
by working out the rate equations which are valid in
the Coulomb blockade regime.\cite{Beenakker:91} Inserting Eq.~(\ref{A_delta_model})
into Eqs.~(\ref{dens_curr_siam}), the reverse-engineering for the Hxc gate
and xc bias potentials can be done analytically. This follows by forming
from Eqs.~(\ref{dens_curr_siam}) the linear combinations $n+I/\g_L$ and
$n-I/\g_R$ and realizing that the inversion of the resulting equations for
the potentials $v\pm V/2$ can be done exactly as in
Refs.~\onlinecite{dittmann:PRL:2018,DittmannHelbigKennes:19}.  
The resulting Hxc gate and xc bias potentials are
\begin{subequations}
\be
\label{v_Hxc}
\tilde{v}_{\rm Hxc}=\frac{1}{2}\left(g(n,-I/\g_R,T_{R})+g(n,I/\g_L,T_{L})\right)
\;,
\ee
\be
\tilde{V}_{\rm xc}=g(n,-I/\g_R,T_{R})-g(n,I/\g_L,T_{L}) \;,
\label{V_xc}
\ee
\label{xcpots_cb}
\end{subequations}
where we have defined
\be
g(n,x,T)= U+T \log\left( \frac{ p+\sqrt{ p^{2}-z y e^{-U/T}}}{y}\right)
 \label{Helbig_functionals}
\ee
with  $z=y-2$, $y=4x+ n$ and $p=n-1+2x\left( 1+ e^{-U/T}\right)$.
As mentioned above, Eqs.~(\ref{xcpots_cb}) are equivalent to
reverse-engineering the rate equations and therefore should be valid at high
temperatures $T\gg\gamma$, i.e. in the parameter regime where the effect of
temperature is much more important than the coupling to the leads.

For the construction of xc potentials which give reasonable approximations
also in the regime of $T\sim\g$ we start by using a model spectral function
\cite{Stefanucci:NL:2015} of the form (\ref{A_delta_model}) but with the
delta functions replaced by Lorentzians
$l_{\gamma}(\omega)=\frac{\gamma}{\omega^{2}+\gamma^{2}/4}$ of width
$\gamma=\g_L+\g_R$, i.e.
\be
A_{\g}^{\rm mod}(\omega)=\left(1-\frac{n}{2}\right)l_{\gamma}(\omega-v)+
\frac{n}{2}l_{\gamma}(\omega-v-U) \;.
\label{A_lorentzian_model}
\ee
\begin{figure}
  \includegraphics[width=\columnwidth]{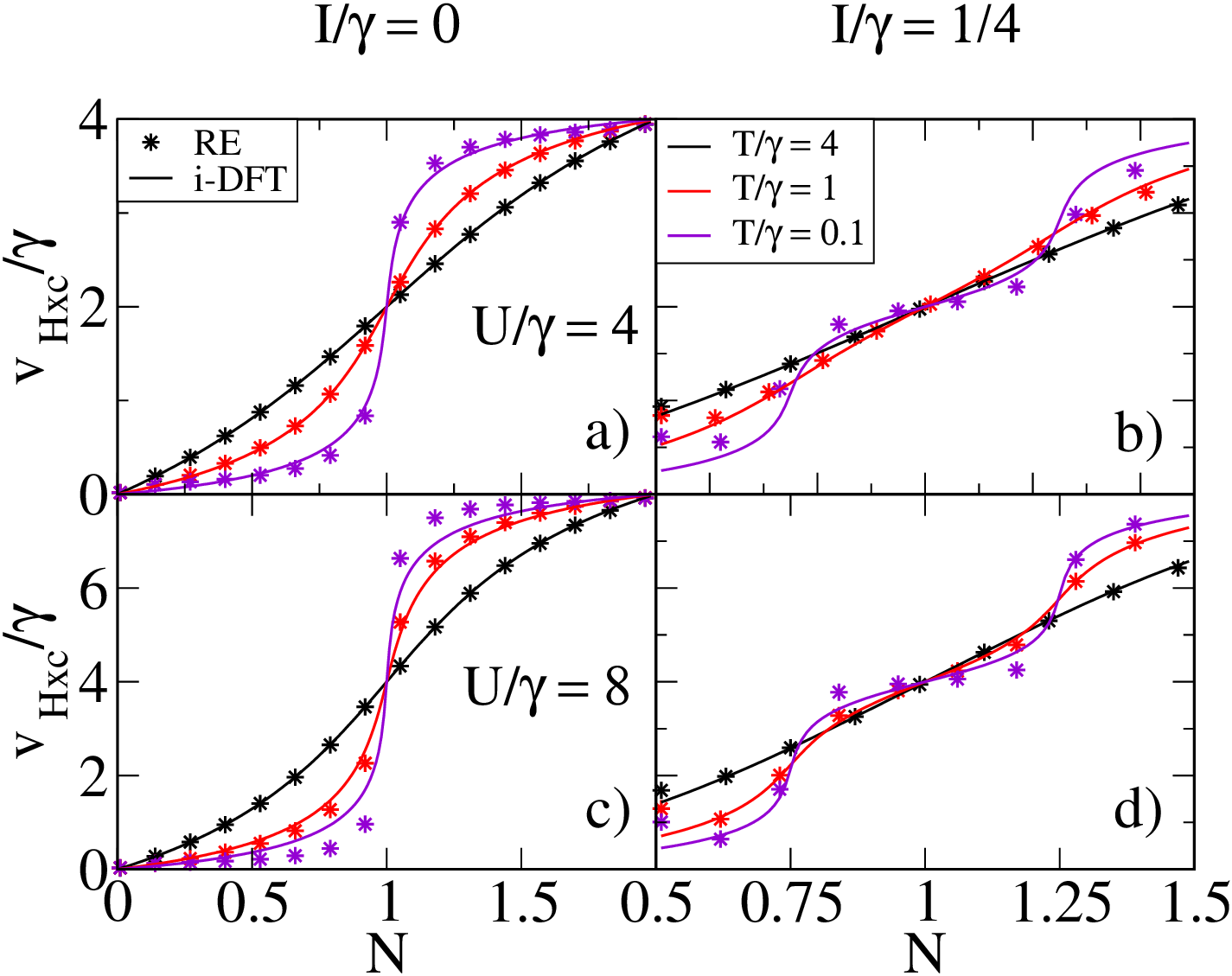}
  \caption{
    Hxc potentials $\tilde{v}_{\rm Hxc}$ of the symmetrically coupled SIAM
    ($\g_L=\g_R=\g/2$) in the Coulomb blockade regime
    for different temperatures calculated by reverse-engineering (RE)
    compared to the parametrization of Eq.~(\ref{xcpots_cb_tempeff})
    Panels a) and b) are for $U/\gamma=4$, panels c) and d) for $U/\gamma=8$,
    while panels a) and c) are for zero current $I=0$
    and panels b) and d) for $I/\g=1/4$. 
  }
  \label{fig:fit_T_star}
\end{figure}
For this model, the reverse-engineering cannot be carried out analytically
(not even at $T=0$) but still can easily be done numerically. However, it is
then desirable to have simple parametrizations of the resulting xc potentials.
In Ref.~\onlinecite{Stefanucci:NL:2015} a simple parametrization for
$T=0$ has been suggested. Since here we are interested in finite temperatures,
a generalization is required. To construct such a parametrization we use the
observation \cite{Kurth:JPCM:2017} that both temperature $T$ and
spectral broadening $\g$ lead to similar smearing out of step features which
are present in the low-temperature and/or strongly correlated limit.
Therefore we suggest a parametrization using the same analytic form as
in Eqs.~(\ref{xcpots_cb}) but replacing the left and right temperatures
$T_{L/R}$ by effective temperatures $T^{*}_{L/R}$, e.g.,
\begin{subequations}
\be
\label{v_Hxc_tempeff}
\tilde{v}_{\rm Hxc}=\frac{1}{2}\left(g(n,-I/\g_R,T^*_{R})+g(n,I/\g_L,T^*_{L})
\right),
\ee
\be
\tilde{V}_{\rm xc}=g(n,-I/\g_R,T^*_{R})-g(n,I/\g_L,T^*_{L}).
\label{V_xc_tempeff}
\ee
\label{xcpots_cb_tempeff}
\end{subequations}
The effective temperatures $T^{*}_{\a}$ we parametrize as 
\be
T^{*}_{\a}(T_{\a},\gamma)=\frac{T_{\a}^{2}+(\eta\gamma)^{2}+\eta\gamma T_{\a}}
{T_{\a}+\eta\gamma} \;.
\label{effective_temperature}
\ee
This parametrization is chosen in such a way that
$T_{\a}^{*}(T_{\a},\gamma\to 0)=T_{\a}$  and
$\eta$ is a fit parameter for which we take the value $\eta=0.45$ in the Coulomb
blockade regime. From now on, we always choose symmetric coupling of the leads,
i.e. $\gamma_L=\gamma_R=\gamma/2$. The quality of our parametrization can be appreciated in
Fig.~\ref{fig:fit_T_star} where we compare the model $\tilde{v}_{\rm Hxc}$
with the corresponding results of the reverse engineering (RE) from Eq.~(\ref{A_lorentzian_model})
 for $U/\g=4$ (panels a) and b)) and $U/\g=8$ (panels c) and d)). We see that, at
equilibrium ($I=0$, panels a) and c)), our parametrization reproduces the
reverse-engineered Hxc potential very accurately for all the considered
temperatures. Also at finite current ($I/\g=1/4$, panels b) and d)), our
approach gives a reasonable parametrization of the reverse-engineered Hxc
potential, although there are some differences at the borders of the domain
for the lowest temperature. 

\begin{figure}
  \includegraphics[width=\linewidth]{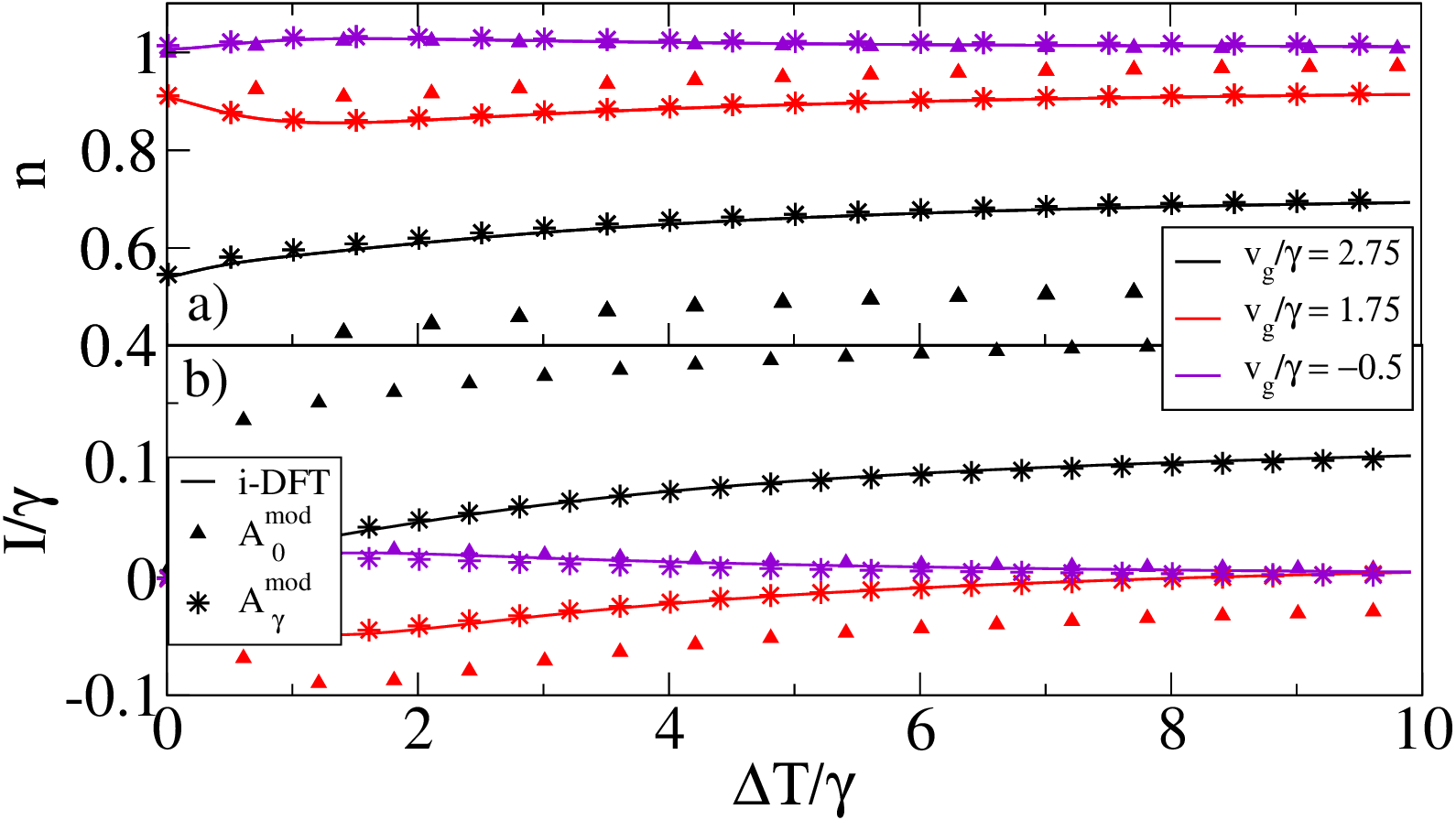}
  \caption{Densities (panel a)) and currents (panel b)) as function of the
    temperature difference $\Delta T = T_{L}-T_{R}$ for $U/\gamma=5$,
    $T_{R}=0.1$ and $V=0$ for different gate voltages $v_g=v+\frac{U}{2}$.
    The i-DFT results using the xc potentials of Eq.~(\ref{xcpots_cb_tempeff})
    are compared with those obtained directly from Eq.~(\ref{dens_curr_siam})
    when using either the model spectral function $A^{mod}_{0}$ of
    Eq.~(\ref{A_delta_model}) or $A^{mod}_{\gamma}$ of
    Eq.~(\ref{A_lorentzian_model}). 
  }
  \label{n_CB}
\end{figure}

As another check on the quality of our parametrization, in Fig.~\ref{n_CB} we
show the density and the electronic current induced by a temperature
difference $\Delta T$ between the leads (at zero bias) as function of
$\Delta T$ for different gate voltages $v_g=v+U/2$. The i-DFT
results are obtained using the xc potentials in Eqs.~(\ref{xcpots_cb_tempeff})
and are compared to those obtained by the direct evaluation of
Eqs.~(\ref{dens_curr_siam}) with the model spectral functions of
Eqs.~(\ref{A_delta_model}) and (\ref{A_lorentzian_model}), respectively. The
agreement, especially for the latter case, is excellent.

An interesting structural property of the xc potentials of
Eqs.~(\ref{xcpots_cb}) is that they are given as the sum of two pieces, each
one depending only on the parameters (temperature and coupling) of one of the
leads. For the couplings this has already been noted in
Ref.~\onlinecite{KurthStefanucci:18}. For the case of different temperatures in
the leads, $T_{\a}=T\pm \Delta T/2$, it can be easily shown that this
structure leads to the xc contribution to the many-body Seebeck coefficient
(Eq.~(\ref{linear_Seebeck_dft})) of the form
\be
S_{xc}^{CB}=\left.\frac{\partial V_{xc}}{\partial \Delta T}\right
\rvert_{\substack{V=0\\ \Delta T=0}}=\left.\frac{\partial v_{Hxc}}{\partial T}
\right\rvert_{\substack{V=0\\ \Delta T=0}}.
\ee
While this result holds for any approximation with the structural property
mentioned above, for the special case of the functionals of
Eq.~(\ref{xcpots_cb}) it reduces exactly to the expression obtained in
Ref.~\onlinecite{Yang:PRB:2016}.\footnote{Notice that by comparing (18) and (9)
  one arrives at $S=S_s-S_{xc}$. This reflects the standard definition $V_{xc}=V_s-V$.}

\begin{figure}
  \includegraphics[width=\linewidth]{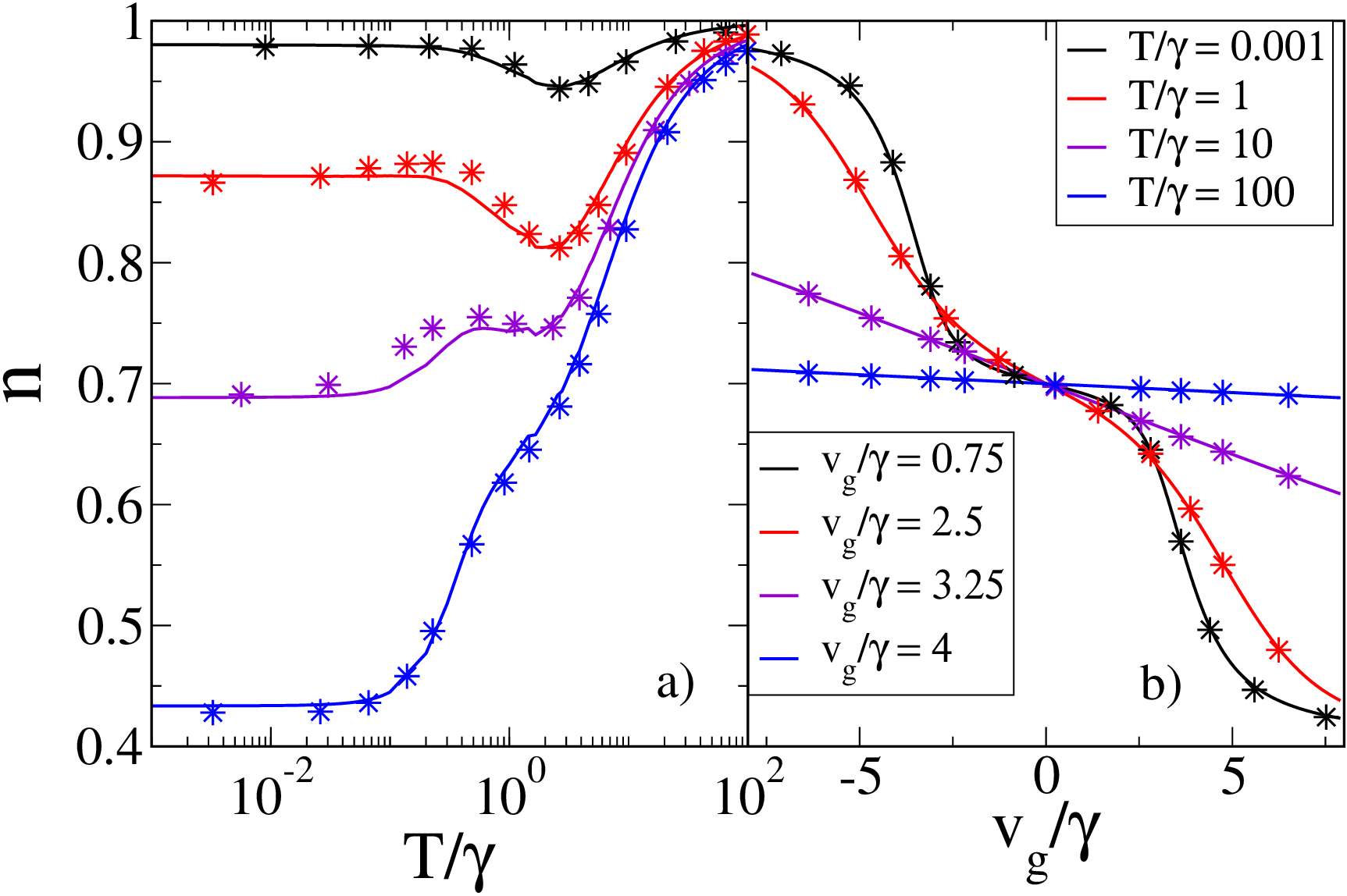}
  \caption{\label{fig:densities}
    Panel a) Densities as function of temperature for different gate
    voltages. Panel b) Densities as function of gate voltage
    for different temperatures. In both panels i-DFT results are compared with
    the NRG results of Ref.~\onlinecite{costi:PRB:2010} for $U/\gamma=8$.
    }
\end{figure}

{\em Kondo regime - }So far we have constructed functionals in the temperature
regime of Coulomb blockade $T\gtrsim T_{K}$ where $T_K$ is the Kondo
temperature. In order to extend the range of applicability of our approximation
to temperatures below $T_K$, we follow the ideas outlined in Ref.~
\onlinecite{Kurth:PRB:2016}. There the central observation was that at zero
temperature the correct behaviour of the zero-bias conductance
(Eq.~(\ref{dI_dV})) is already contained in the KS conductance $G_s$ due to
the Friedel sum rule.
\cite{StefanucciKurth:11,Bergfield:PRL:2012,TroesterSchmitteckertEvers:12}
Therefore, at zero temperature the derivative $\partial V_{\rm xc}/\partial I$
has to vanish at $I=0$. Following Ref.~\onlinecite{Kurth:PRB:2016}, we modify
our functional as
\begin{subequations}
\be
v_{Hxc}=\left[1-k(n,I,T)\right]\tilde{v}_{Hxc}+k(n,I,T)v_{Hxc}^{(0)}(n)
\ee
\be
V_{xc}=\left[1-k(n,I,T)\right]\tilde{V}_{xc}(n,I,T)
\ee
\label{functionals_prefactor}
\end{subequations}
where $v_{Hxc}^{(0)}$ is the zero-temperature, equilibrium Hxc potential of
Ref.~\onlinecite{Bergfield:PRL:2012} which accurately parametrizes density
matrix renormalization group results. We further introduce the prefactor
$k(n,I,T)$ with the properties 
$k(n,I=0,T=0)=1$ and $\partial k(n,I,T)/\partial I \vert_{I=0,T=0}=0$.
The first property ensures that at zero current and zero temperature
$v_{\rm Hxc}$ reduces to $v_{\rm Hxc}^{(0)}$, the second one leads to a vanishing
correction to the KS zero-bias conductance at zero temperature.
To be specific, we choose $k(n,I,T)=a(n,I)z(T)$, where $a(n,I)$ is the same
prefactor used in Eq.~(12) of Ref.~\onlinecite{Kurth:PRB:2016}. This
prefactor, although combined with a different form for the Coulomb blockade
functionals $\tilde{v}_{\rm Hxc}$ and $\tilde{V}_{\rm xc}$ ensures
a good description of the finite bias conductance for relatively low
temperatures. We also found it convenient to introduce another prefactor
$z(T)=(1+(2.5T/\gamma)^{3})^{-1}$ to ensure a smooth transition to the Coulomb
blockade form of the functional at high temperatures.
Finally, we redefine $\eta=0.1U/\gamma + 0.36$ entering in the effective
temperature $T^*$ of Eq.~(\ref{effective_temperature}) in order to correct the effect of the interactions at low temperatures.
This is somewhat similar to
Ref.~\onlinecite{Kurth:PRB:2016} where the smoothening of the step features
in the Coulomb blockade part of the functional had to be modified in the
Kondo regime $T\lesssim T_K$. 

As a first test of this functional, we calculate self-consistent densities
at equilibrium. In Fig.~\ref{fig:densities} a) we plot the densities obtained
for different gate voltages as function of the temperature of
the leads $T=T_{L}=T_{R}$ for the strongly correlated case with $U/\gamma=8$
and compare with numerical renormalization group (NRG) results of
Ref.~\onlinecite{costi:PRB:2010}. Instead, in Fig.~\ref{fig:densities} b) we
show equilibrium densities as function of gate voltage for different
temperatures. The agreement of our i-DFT densities with the NRG ones is
excellent. 

\begin{figure}
  \includegraphics[width=\linewidth]{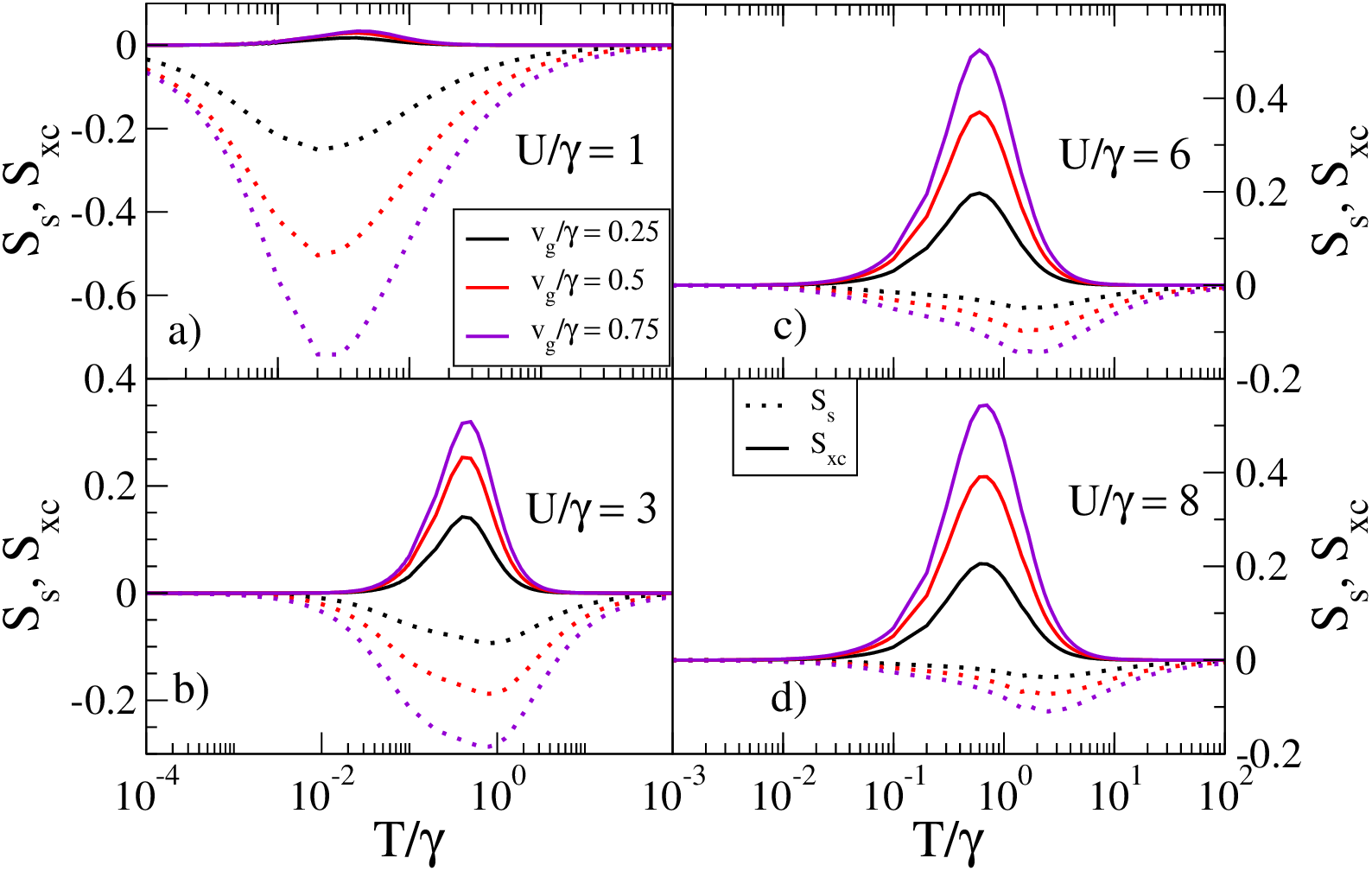}
  \caption{Linear KS Seebeck coefficient $S_{s}$ and xc correction $S_{\rm xc}$
    as function of temperature for different gate voltages and
    correlation strengths $U/\gamma$. }
  \label{fig:S_s_and_S_xc}
\end{figure}

\begin{figure}
  \includegraphics[width=\linewidth]{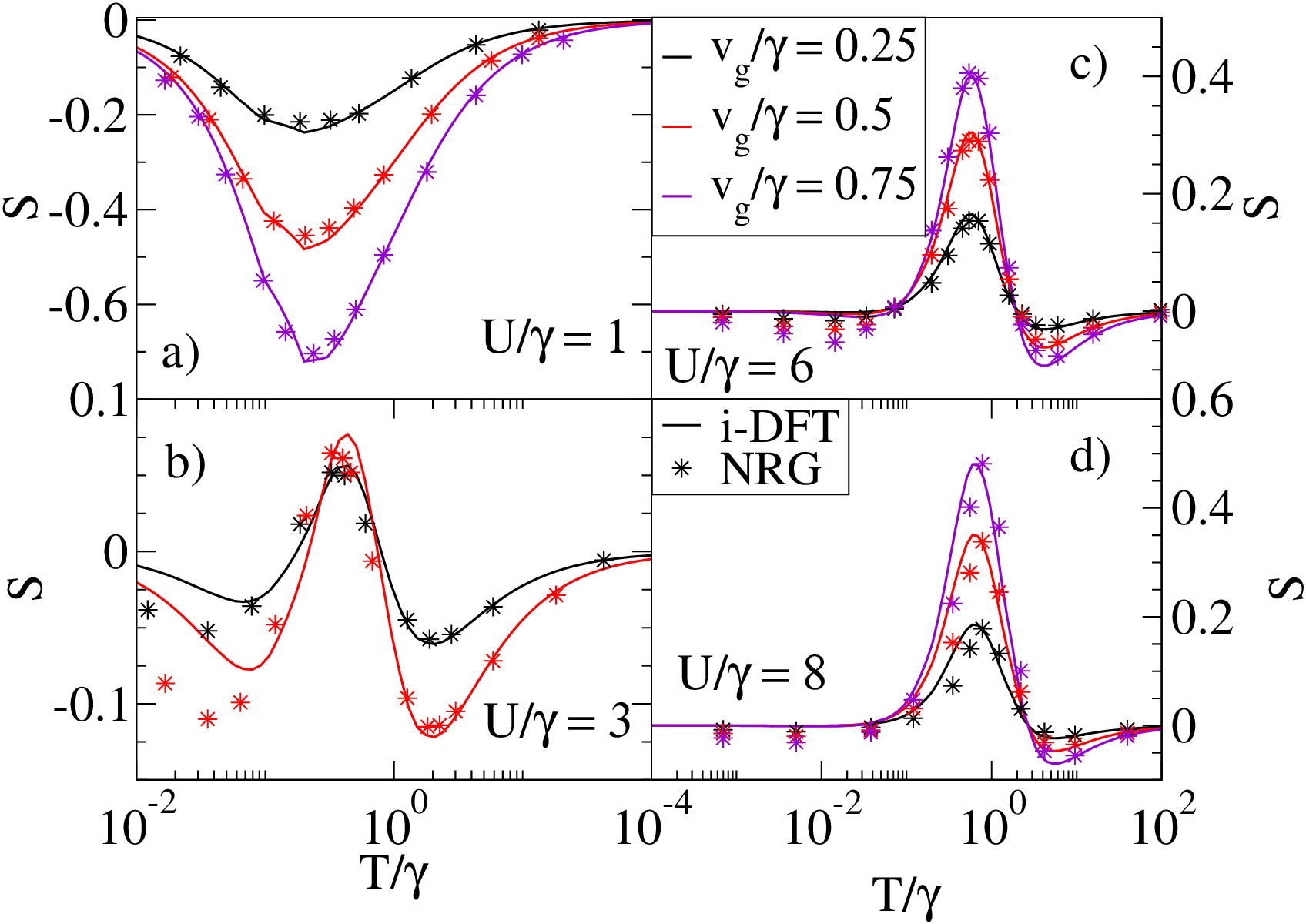}
  \caption{Comparison of the Seebeck coefficient obtained with i-DFT with the
    NRG results of Ref.~\onlinecite{costi:PRB:2010} as function of
    temperature for different gate voltages and correlation
    strengths $U/\gamma$.
  }
  \label{fig:S_K_T}
\end{figure}

We now turn to the Seebeck coefficient. As first step we analyze the relative
magnitude of the KS Seebeck coefficient ($S_s$) and the xc correction
(see Eq.~\ref{linear_Seebeck_dft}) as a function of the gate voltage and
the correlation strength. In Fig.~\ref{fig:S_s_and_S_xc}, we can
see that, as expected, the xc contribution becomes dominant for almost any
temperature as $U/\gamma$ increases from 1 (Fig.~\ref{fig:S_s_and_S_xc}a)
to 8 (Fig.~\ref{fig:S_s_and_S_xc}d), but also for intermediate values,
$U/\gamma=3$, the two terms have a comparable magnitude for any value of the
gate voltage. Notice that, since both potential and temperature are evaluated
in units of $\gamma$, $S$ and $S_{xc}$ are dimensionless.
In Fig.~\ref{fig:S_K_T} we compare
our results with the NRG ones of Ref.~\onlinecite{costi:PRB:2010} for fixed
gate potential as a function of temperature. Again, similar to
Fig.~\ref{fig:S_s_and_S_xc}, the panels report calculated values from weak (a) to
strong correlations (d) in the dot. As expected we find a very good agreement
between i-DFT and NRG for $T\gtrsim T_{K}$. For lower temperatures, i-DFT
shows small discrepancies with respect to the reference result which
exhibits a different evolution of the local minimum of the Seebeck coefficient
when increasing the interaction.

\begin{figure}
  \includegraphics[width=0.9\linewidth]{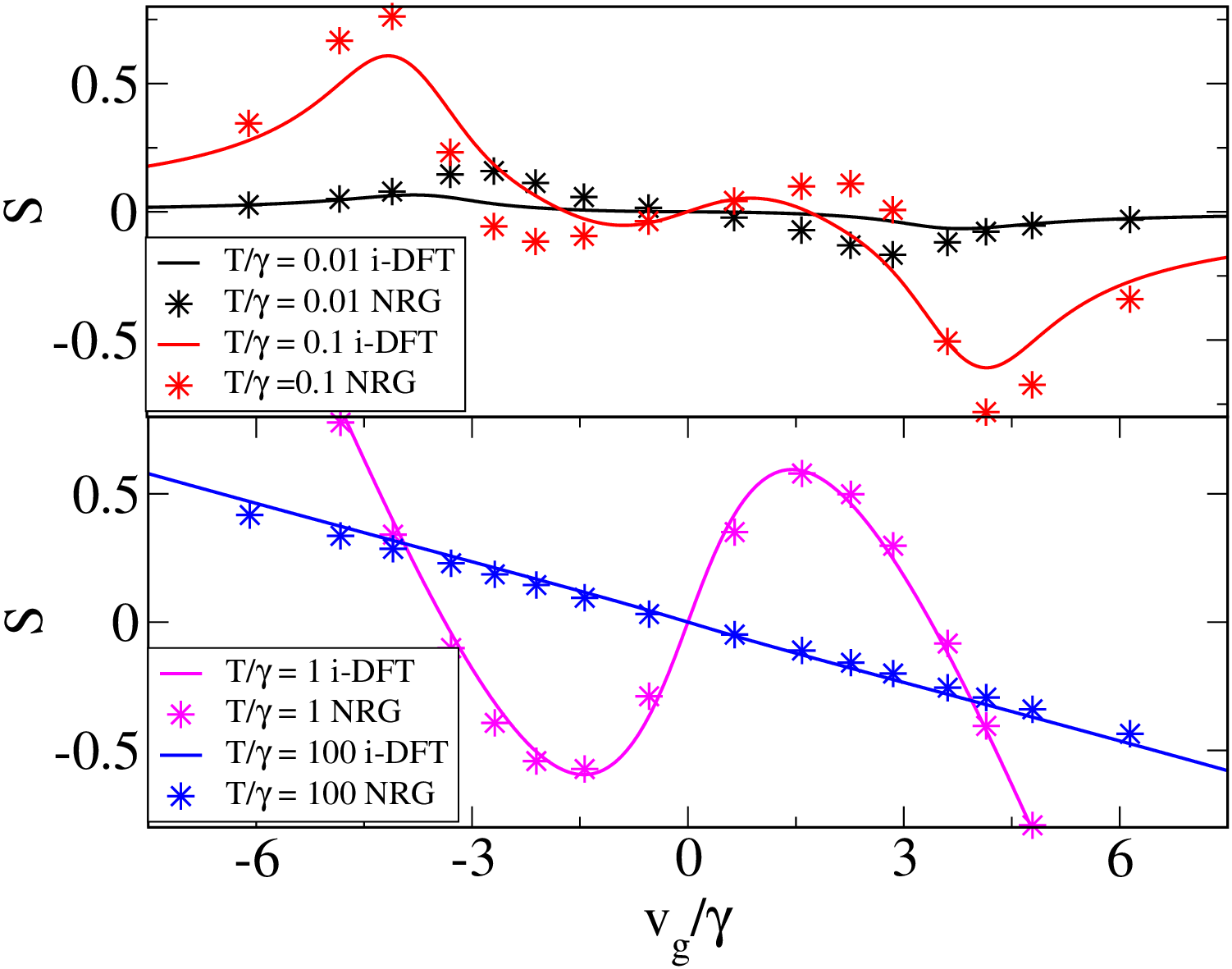}
  \caption{Comparison of the Seebeck coefficient obtained with i-DFT with the
    NRG results of Ref.~\onlinecite{costi:PRB:2010}.$S$ is shown as function
    of the gate voltage for different temperatures and
    for strong correlations $U/\gamma=8$. 
  }
  \label{fig:S_K_vg}
\end{figure}

In Fig.~\ref{fig:S_K_vg}, we show the Seebeck coefficient as function of the
gate voltage for different values of the temperature and again compare with
NRG results of Ref.~\onlinecite{costi:PRB:2010}. As already noticed above, for
low temperatures there are discrepancies at certain gate values although with
our i-DFT approach we manage to obtain the qualitative behaviour of the NRG
results. For $T/\g \gtrsim1$, on the other hand, the i-DFT results are in
excellent agreement with the NRG ones.

\begin{figure}
  \includegraphics[width=0.9\linewidth]{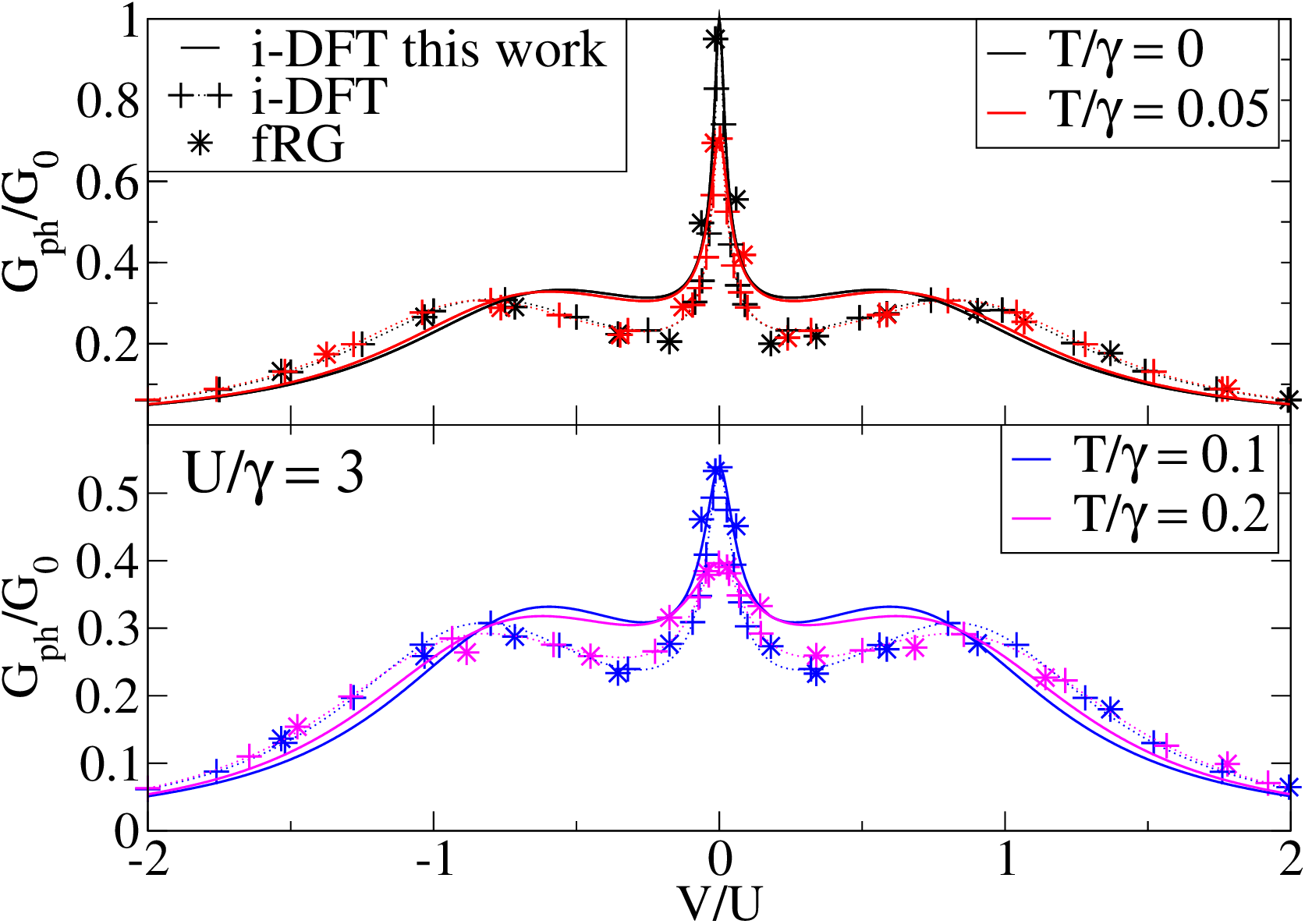}
  \caption{Differential conductance $G_{\rm ph}$ at the particle hole symmetric
    point $v=-U/2$ for the SIAM as function of bias $V$ for
    $U/\gamma=3$. The i-DFT results obtained with the functional of
    Eq.~(\ref{functionals_prefactor}) are compared to those from
    Ref.~\onlinecite{Kurth:PRB:2016} and the fRG results of
    Ref.~\onlinecite{Jakobs:PRB:2010}. $G_{0}=1/\pi$ is the quantum of
    conductance.   }
  \label{fig:conductance_ph}
\end{figure}

\begin{figure}
  \includegraphics[width=\linewidth]{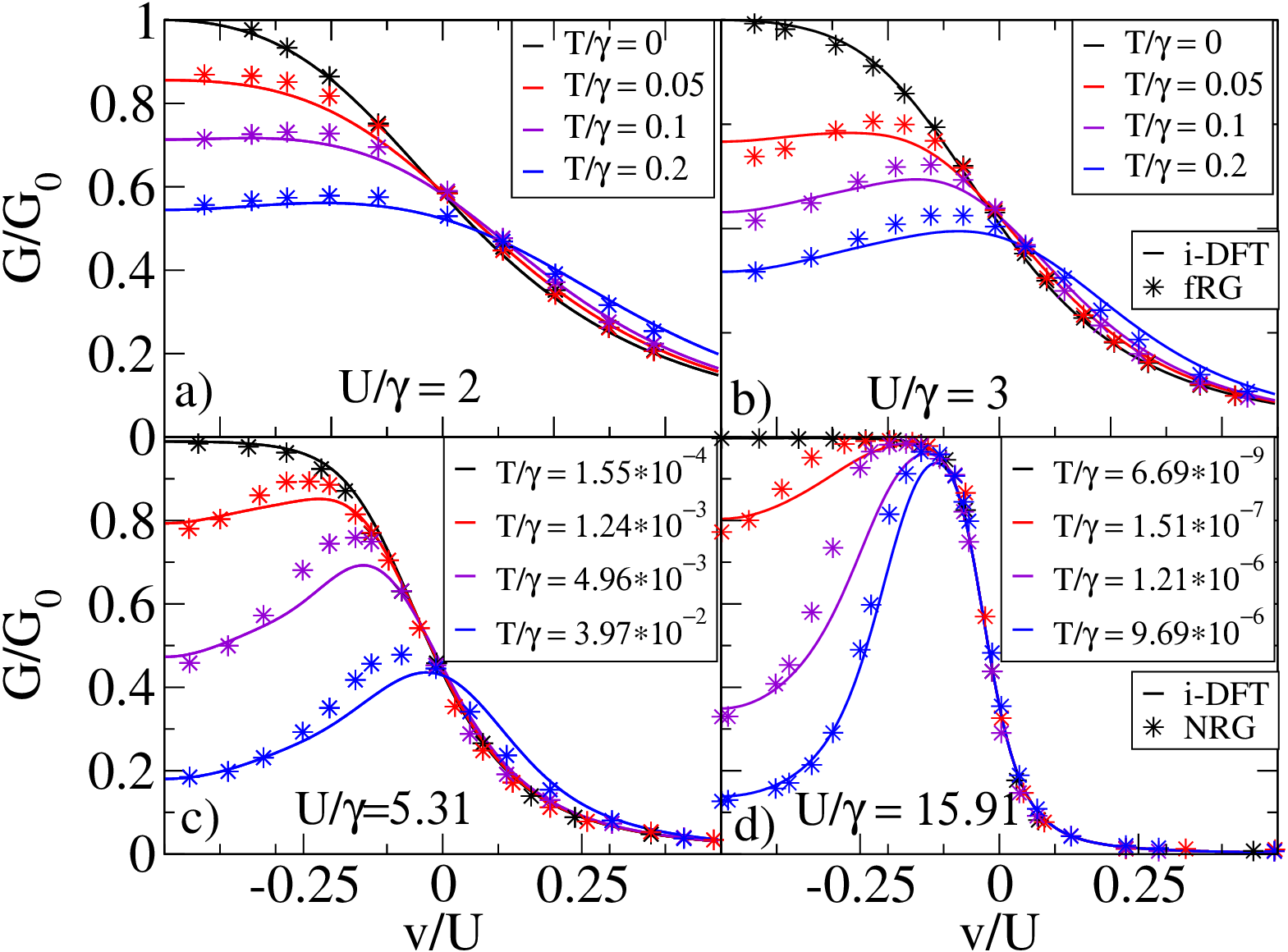}
  \caption{\label{fig:conductance_zb}
    Zero-bias differential conductance of i-DFT obtained
    with the functional of Eq.~(\ref{functionals_prefactor}) as function of
    the gate level $v$ for different values of $U/\gamma$. In panels
    a) and b) the results are compared to fRG ones from
    Ref.~\onlinecite{Jakobs:PRB:2010}, in panels c) and d) NRG results from
    Ref.~\onlinecite{Izumida:PSJ:2001} are used for comparison.  $G_{0}=1/\pi$
    is the quantum of conductance.
    }
\end{figure}

While the main focus of the present work is on the Seebeck coefficient, one
can, of course, also calculate differential conductances from i-DFT. In
Fig.~\ref{fig:conductance_ph}, we show differential conductances for the SIAM
at the particle-hole symmetric point obtained with our present functional and
compare them with those obtained using the functional of
Ref.~\onlinecite{Kurth:PRB:2016} as well as with functional renormalization
group (fRG) results of Ref.~\onlinecite{Jakobs:PRB:2010}. The i-DFT results
with our present functional agree reasonably well with the reference fRG results
although some details like the overall shape of the side peaks seem to be
better captured by the functional of Ref.~\onlinecite{Kurth:PRB:2016}. 
Finally, the differential conductance at zero bias has been calculated and
compared with both fRG of Ref.~\onlinecite{Jakobs:PRB:2010} and NRG of
Ref.~\onlinecite{Izumida:PSJ:2001} for different interaction
strengths obtaining very good agreements, as can be appreciated in
Fig.~\ref{fig:conductance_zb}.

\section{Conclusions}

In this work we have extended the recently proposed density functional
framework for steady-state transport (i-DFT) to the situation when there is a
temperature gradient across the junction. As a direct consequence, we have
derived an exact expression for the Seebeck coefficient of the interacting
system as a sum of the KS Seebeck coefficient and a correction term related
to a derivative of the i-DFT xc bias functional. 

For the SIAM we have constructed an approximation to the (H)xc functionals
both in the Coulomb blockade as well as in the Kondo regime. In the Coulomb
blockade regime we found that both Hxc gate and xc bias potential consist
of a sum or difference of two pieces, each of which depends only on the
temperature of one of the leads. This property allows us to recover an earlier,
approximate expression for the Seebeck coefficient in terms of purely DFT
quantities\cite{Yang:PRB:2016}.
We have compared the Seebeck coefficients for the SIAM obtained with our method
with those from NRG calculations reported in the literature. While our
parametrization by construction becomes exact at high temperatures
($T\gg T_K$), in the Kondo regime ($T\ll T_K$) the agreement is reasonable but
not perfect. However, i-DFT is numerically much cheaper than more
sophisticated many-body methods and extension to more complicated models
can also be relatively straightforward. 

\acknowledgments
We gratefully acknowledge useful discussions with Florian Eich, David Jacob
and Gianluca Stefanucci. 
We acknowledge funding by the grant ``Grupos Consolidados UPV/EHU del Gobierno
Vasco'' (IT1249-19) as well as the grant of the ``Ministerio de Economia y
Competividad (MINECO)'' (FIS2016-79464-P).


\begin{thebibliography}{41}%
\makeatletter
\providecommand \@ifxundefined [1]{%
 \@ifx{#1\undefined}
}%
\providecommand \@ifnum [1]{%
 \ifnum #1\expandafter \@firstoftwo
 \else \expandafter \@secondoftwo
 \fi
}%
\providecommand \@ifx [1]{%
 \ifx #1\expandafter \@firstoftwo
 \else \expandafter \@secondoftwo
 \fi
}%
\providecommand \natexlab [1]{#1}%
\providecommand \enquote  [1]{``#1''}%
\providecommand \bibnamefont  [1]{#1}%
\providecommand \bibfnamefont [1]{#1}%
\providecommand \citenamefont [1]{#1}%
\providecommand \href@noop [0]{\@secondoftwo}%
\providecommand \href [0]{\begingroup \@sanitize@url \@href}%
\providecommand \@href[1]{\@@startlink{#1}\@@href}%
\providecommand \@@href[1]{\endgroup#1\@@endlink}%
\providecommand \@sanitize@url [0]{\catcode `\\12\catcode `\$12\catcode
  `\&12\catcode `\#12\catcode `\^12\catcode `\_12\catcode `\%12\relax}%
\providecommand \@@startlink[1]{}%
\providecommand \@@endlink[0]{}%
\providecommand \url  [0]{\begingroup\@sanitize@url \@url }%
\providecommand \@url [1]{\endgroup\@href {#1}{\urlprefix }}%
\providecommand \urlprefix  [0]{URL }%
\providecommand \Eprint [0]{\href }%
\providecommand \doibase [0]{http://dx.doi.org/}%
\providecommand \selectlanguage [0]{\@gobble}%
\providecommand \bibinfo  [0]{\@secondoftwo}%
\providecommand \bibfield  [0]{\@secondoftwo}%
\providecommand \translation [1]{[#1]}%
\providecommand \BibitemOpen [0]{}%
\providecommand \bibitemStop [0]{}%
\providecommand \bibitemNoStop [0]{.\EOS\space}%
\providecommand \EOS [0]{\spacefactor3000\relax}%
\providecommand \BibitemShut  [1]{\csname bibitem#1\endcsname}%
\let\auto@bib@innerbib\@empty
\bibitem [{\citenamefont {Vining}(2008)}]{Vining:NAT:2008}%
  \BibitemOpen
  \bibfield  {author} {\bibinfo {author} {\bibfnamefont {C.~B.}\ \bibnamefont
  {Vining}},\ }\href@noop {} {\bibfield  {journal} {\bibinfo  {journal} {Nature
  Mat.}\ }\textbf {\bibinfo {volume} {7}},\ \bibinfo {pages} {765} (\bibinfo
  {year} {2008})}\BibitemShut {NoStop}%
\bibitem [{\citenamefont {Vining}(2009)}]{Vining:NAT:2009}%
  \BibitemOpen
  \bibfield  {author} {\bibinfo {author} {\bibfnamefont {C.~B.}\ \bibnamefont
  {Vining}},\ }\href@noop {} {\bibfield  {journal} {\bibinfo  {journal} {Nature
  Mat.}\ }\textbf {\bibinfo {volume} {8}},\ \bibinfo {pages} {83} (\bibinfo
  {year} {2009})}\BibitemShut {NoStop}%
\bibitem [{\citenamefont {Dubi}\ and\ \citenamefont
  {Di~Ventra}(2011)}]{Dubi:RMP:2011}%
  \BibitemOpen
  \bibfield  {author} {\bibinfo {author} {\bibfnamefont {Y.}~\bibnamefont
  {Dubi}}\ and\ \bibinfo {author} {\bibfnamefont {M.}~\bibnamefont
  {Di~Ventra}},\ }\href@noop {} {\bibfield  {journal} {\bibinfo  {journal}
  {Rev. Mod. Phys.}\ }\textbf {\bibinfo {volume} {83}},\ \bibinfo {pages} {131}
  (\bibinfo {year} {2011})}\BibitemShut {NoStop}%
\bibitem [{\citenamefont {Goldsmid}(2010)}]{goldsmid:2010:introduction}%
  \BibitemOpen
  \bibfield  {author} {\bibinfo {author} {\bibfnamefont {H.~J.}\ \bibnamefont
  {Goldsmid}},\ }\href@noop {} {\emph {\bibinfo {title} {Introduction to
  thermoelectricity}}},\ Vol.\ \bibinfo {volume} {121}\ (\bibinfo  {publisher}
  {Springer},\ \bibinfo {year} {2010})\BibitemShut {NoStop}%
\bibitem [{\citenamefont {Hicks}\ and\ \citenamefont
  {Dresselhaus}(1993)}]{Hicks1993}%
  \BibitemOpen
  \bibfield  {author} {\bibinfo {author} {\bibfnamefont {L.~D.}\ \bibnamefont
  {Hicks}}\ and\ \bibinfo {author} {\bibfnamefont {M.~S.}\ \bibnamefont
  {Dresselhaus}},\ }\href {\doibase 10.1103/PhysRevB.47.12727} {\bibfield
  {journal} {\bibinfo  {journal} {Phys. Rev. B}\ }\textbf {\bibinfo {volume}
  {47}},\ \bibinfo {pages} {12727} (\bibinfo {year} {1993})}\BibitemShut
  {NoStop}%
\bibitem [{\citenamefont {Datta}(1997)}]{Datta1997}%
  \BibitemOpen
  \bibfield  {author} {\bibinfo {author} {\bibfnamefont {S.}~\bibnamefont
  {Datta}},\ }\href@noop {} {\emph {\bibinfo {title} {{Electronic Transport in
  Mesoscopic Systems}}}}\ (\bibinfo  {publisher} {Cambridge University Press},\
  \bibinfo {address} {New York},\ \bibinfo {year} {1997})\BibitemShut {NoStop}%
\bibitem [{\citenamefont {Cuniberti}\ \emph {et~al.}(2005)\citenamefont
  {Cuniberti}, \citenamefont {Fagas},\ and\ \citenamefont
  {Richter}}]{Cuniberti2005}%
  \BibitemOpen
  \bibinfo {editor} {\bibfnamefont {G.}~\bibnamefont {Cuniberti}}, \bibinfo
  {editor} {\bibfnamefont {G.}~\bibnamefont {Fagas}}, \ and\ \bibinfo {editor}
  {\bibfnamefont {K.}~\bibnamefont {Richter}},\ eds.,\ \href@noop {} {\emph
  {\bibinfo {title} {{Introducing Molecular Electronics}}}}\ (\bibinfo
  {publisher} {Springer},\ \bibinfo {address} {New York},\ \bibinfo {year}
  {2005})\BibitemShut {NoStop}%
\bibitem [{\citenamefont {Cuevas}\ and\ \citenamefont
  {Scheer}(2010)}]{CuevasScheer:10}%
  \BibitemOpen
  \bibfield  {author} {\bibinfo {author} {\bibfnamefont {J.~C.}\ \bibnamefont
  {Cuevas}}\ and\ \bibinfo {author} {\bibfnamefont {E.}~\bibnamefont
  {Scheer}},\ }\href@noop {} {\emph {\bibinfo {title} {Molecular Electronics:
  An Introduction to Theory and Experiment}}}\ (\bibinfo  {publisher} {World
  Scientific},\ \bibinfo {address} {London},\ \bibinfo {year}
  {2010})\BibitemShut {NoStop}%
\bibitem [{\citenamefont {Beenakker}(1991)}]{Beenakker:91}%
  \BibitemOpen
  \bibfield  {author} {\bibinfo {author} {\bibfnamefont {C.~W.~J.}\
  \bibnamefont {Beenakker}},\ }\href {\doibase 10.1103/PhysRevB.44.1646}
  {\bibfield  {journal} {\bibinfo  {journal} {Phys. Rev. B}\ }\textbf {\bibinfo
  {volume} {44}},\ \bibinfo {pages} {1646} (\bibinfo {year}
  {1991})}\BibitemShut {NoStop}%
\bibitem [{\citenamefont {Zianni}(2008)}]{Zianni2008}%
  \BibitemOpen
  \bibfield  {author} {\bibinfo {author} {\bibfnamefont {X.}~\bibnamefont
  {Zianni}},\ }\href {\doibase 10.1103/PhysRevB.78.165327} {\bibfield
  {journal} {\bibinfo  {journal} {Phys. Rev. B}\ }\textbf {\bibinfo {volume}
  {78}},\ \bibinfo {pages} {165327} (\bibinfo {year} {2008})}\BibitemShut
  {NoStop}%
\bibitem [{\citenamefont {Sothmann}\ \emph {et~al.}(2013)\citenamefont
  {Sothmann}, \citenamefont {S{\'{a}}nchez}, \citenamefont {Jordan},\ and\
  \citenamefont {B{\"{u}}ttiker}}]{Sothmann2013}%
  \BibitemOpen
  \bibfield  {author} {\bibinfo {author} {\bibfnamefont {B.}~\bibnamefont
  {Sothmann}}, \bibinfo {author} {\bibfnamefont {R.}~\bibnamefont
  {S{\'{a}}nchez}}, \bibinfo {author} {\bibfnamefont {A.~N.}\ \bibnamefont
  {Jordan}}, \ and\ \bibinfo {author} {\bibfnamefont {M.}~\bibnamefont
  {B{\"{u}}ttiker}},\ }\href {\doibase 10.1088/1367-2630/15/9/095021}
  {\bibfield  {journal} {\bibinfo  {journal} {New J. Phys.}\ }\textbf {\bibinfo
  {volume} {15}},\ \bibinfo {pages} {095021} (\bibinfo {year}
  {2013})}\BibitemShut {NoStop}%
\bibitem [{\citenamefont {S{\'{a}}nchez}\ \emph {et~al.}(2013)\citenamefont
  {S{\'{a}}nchez}, \citenamefont {Sothmann}, \citenamefont {Jordan},\ and\
  \citenamefont {B{\"{u}}ttiker}}]{Sanchez2013a}%
  \BibitemOpen
  \bibfield  {author} {\bibinfo {author} {\bibfnamefont {R.}~\bibnamefont
  {S{\'{a}}nchez}}, \bibinfo {author} {\bibfnamefont {B.}~\bibnamefont
  {Sothmann}}, \bibinfo {author} {\bibfnamefont {A.~N.}\ \bibnamefont
  {Jordan}}, \ and\ \bibinfo {author} {\bibfnamefont {M.}~\bibnamefont
  {B{\"{u}}ttiker}},\ }\href {\doibase 10.1088/1367-2630/15/12/125001}
  {\bibfield  {journal} {\bibinfo  {journal} {New J. Phys.}\ }\textbf {\bibinfo
  {volume} {15}},\ \bibinfo {pages} {125001} (\bibinfo {year}
  {2013})}\BibitemShut {NoStop}%
\bibitem [{\citenamefont {D'Agosta}(2013)}]{DAgosta2013}%
  \BibitemOpen
  \bibfield  {author} {\bibinfo {author} {\bibfnamefont {R.}~\bibnamefont
  {D'Agosta}},\ }\href {\doibase 10.1039/c2cp42594g} {\bibfield  {journal}
  {\bibinfo  {journal} {Phys. Chem. Chem. Phys.}\ }\textbf {\bibinfo {volume}
  {15}},\ \bibinfo {pages} {1758} (\bibinfo {year} {2013})}\BibitemShut
  {NoStop}%
\bibitem [{\citenamefont {Eich}\ \emph {et~al.}(2014)\citenamefont {Eich},
  \citenamefont {Principi}, \citenamefont {{Di Ventra}},\ and\ \citenamefont
  {Vignale}}]{Eich2014a}%
  \BibitemOpen
  \bibfield  {author} {\bibinfo {author} {\bibfnamefont {F.~G.}\ \bibnamefont
  {Eich}}, \bibinfo {author} {\bibfnamefont {A.}~\bibnamefont {Principi}},
  \bibinfo {author} {\bibfnamefont {M.}~\bibnamefont {{Di Ventra}}}, \ and\
  \bibinfo {author} {\bibfnamefont {G.}~\bibnamefont {Vignale}},\ }\href
  {\doibase 10.1103/PhysRevB.90.115116} {\bibfield  {journal} {\bibinfo
  {journal} {Phys. Rev. B}\ }\textbf {\bibinfo {volume} {90}},\ \bibinfo
  {pages} {115116} (\bibinfo {year} {2014})}\BibitemShut {NoStop}%
\bibitem [{\citenamefont {Hohenberg}\ and\ \citenamefont
  {Kohn}(1964)}]{Hohenberg1964}%
  \BibitemOpen
  \bibfield  {author} {\bibinfo {author} {\bibfnamefont {P.}~\bibnamefont
  {Hohenberg}}\ and\ \bibinfo {author} {\bibfnamefont {W.}~\bibnamefont
  {Kohn}},\ }\href {\doibase 10.1103/PhysRev.136.B864} {\bibfield  {journal}
  {\bibinfo  {journal} {Phys. Rev.}\ }\textbf {\bibinfo {volume} {136}},\
  \bibinfo {pages} {B864} (\bibinfo {year} {1964})}\BibitemShut {NoStop}%
\bibitem [{\citenamefont {Kohn}\ and\ \citenamefont {Sham}(1965)}]{Kohn1965}%
  \BibitemOpen
  \bibfield  {author} {\bibinfo {author} {\bibfnamefont {W.}~\bibnamefont
  {Kohn}}\ and\ \bibinfo {author} {\bibfnamefont {L.~J.}\ \bibnamefont
  {Sham}},\ }\href {\doibase 10.1103/PhysRev.140.A1133} {\bibfield  {journal}
  {\bibinfo  {journal} {Phys. Rev.}\ }\textbf {\bibinfo {volume} {140}},\
  \bibinfo {pages} {A1133} (\bibinfo {year} {1965})}\BibitemShut {NoStop}%
\bibitem [{\citenamefont {Thoss}\ and\ \citenamefont
  {Evers}(2018)}]{ThossEvers:18}%
  \BibitemOpen
  \bibfield  {author} {\bibinfo {author} {\bibfnamefont {M.}~\bibnamefont
  {Thoss}}\ and\ \bibinfo {author} {\bibfnamefont {F.}~\bibnamefont {Evers}},\
  }\href {https://doi.org/10.1063/1.5003306} {\bibfield  {journal} {\bibinfo
  {journal} {J. Chem. Phys.}\ }\textbf {\bibinfo {volume} {148}},\ \bibinfo
  {pages} {030901} (\bibinfo {year} {2018})}\BibitemShut {NoStop}%
\bibitem [{\citenamefont {Runge}\ and\ \citenamefont {{E.K.U.
  Gross}}(1984)}]{RungeGross:84}%
  \BibitemOpen
  \bibfield  {author} {\bibinfo {author} {\bibfnamefont {E.}~\bibnamefont
  {Runge}}\ and\ \bibinfo {author} {\bibnamefont {{E.K.U. Gross}}},\
  }\href@noop {} {\bibfield  {journal} {\bibinfo  {journal} {Phys. Rev. Lett}\
  }\textbf {\bibinfo {volume} {52}},\ \bibinfo {pages} {997} (\bibinfo {year}
  {1984})}\BibitemShut {NoStop}%
\bibitem [{\citenamefont {Stefanucci}\ and\ \citenamefont
  {{C.-O.~Almbladh}}(2004{\natexlab{a}})}]{StefanucciAlmbladh:04}%
  \BibitemOpen
  \bibfield  {author} {\bibinfo {author} {\bibfnamefont {G.}~\bibnamefont
  {Stefanucci}}\ and\ \bibinfo {author} {\bibnamefont {{C.-O.~Almbladh}}},\
  }\href@noop {} {\bibfield  {journal} {\bibinfo  {journal} {EPL}\ }\textbf
  {\bibinfo {volume} {67}},\ \bibinfo {pages} {14} (\bibinfo {year}
  {2004}{\natexlab{a}})}\BibitemShut {NoStop}%
\bibitem [{\citenamefont {Stefanucci}\ and\ \citenamefont
  {{C.-O.~Almbladh}}(2004{\natexlab{b}})}]{StefanucciAlmbladh:04-2}%
  \BibitemOpen
  \bibfield  {author} {\bibinfo {author} {\bibfnamefont {G.}~\bibnamefont
  {Stefanucci}}\ and\ \bibinfo {author} {\bibnamefont {{C.-O.~Almbladh}}},\
  }\href@noop {} {\bibfield  {journal} {\bibinfo  {journal} {Phys. Rev. B}\
  }\textbf {\bibinfo {volume} {69}},\ \bibinfo {pages} {195318} (\bibinfo
  {year} {2004}{\natexlab{b}})}\BibitemShut {NoStop}%
\bibitem [{\citenamefont {Sai}\ \emph {et~al.}(2005)\citenamefont {Sai},
  \citenamefont {Zwolak}, \citenamefont {Vignale},\ and\ \citenamefont
  {Di~Ventra}}]{SaiZwolakVignaleDiVentra:05}%
  \BibitemOpen
  \bibfield  {author} {\bibinfo {author} {\bibfnamefont {N.}~\bibnamefont
  {Sai}}, \bibinfo {author} {\bibfnamefont {M.}~\bibnamefont {Zwolak}},
  \bibinfo {author} {\bibfnamefont {G.}~\bibnamefont {Vignale}}, \ and\
  \bibinfo {author} {\bibfnamefont {M.}~\bibnamefont {Di~Ventra}},\ }\href
  {\doibase 10.1103/PhysRevLett.94.186810} {\bibfield  {journal} {\bibinfo
  {journal} {Phys. Rev. Lett.}\ }\textbf {\bibinfo {volume} {94}},\ \bibinfo
  {pages} {186810} (\bibinfo {year} {2005})}\BibitemShut {NoStop}%
\bibitem [{\citenamefont {Koentopp}\ \emph {et~al.}(2006)\citenamefont
  {Koentopp}, \citenamefont {Burke},\ and\ \citenamefont
  {Evers}}]{KoentoppBurkeEvers:06}%
  \BibitemOpen
  \bibfield  {author} {\bibinfo {author} {\bibfnamefont {M.}~\bibnamefont
  {Koentopp}}, \bibinfo {author} {\bibfnamefont {K.}~\bibnamefont {Burke}}, \
  and\ \bibinfo {author} {\bibfnamefont {F.}~\bibnamefont {Evers}},\ }\href
  {\doibase 10.1103/PhysRevB.73.121403} {\bibfield  {journal} {\bibinfo
  {journal} {Phys. Rev. B}\ }\textbf {\bibinfo {volume} {73}},\ \bibinfo
  {pages} {121403} (\bibinfo {year} {2006})}\BibitemShut {NoStop}%
\bibitem [{\citenamefont {Stefanucci}\ \emph {et~al.}(2006)\citenamefont
  {Stefanucci}, \citenamefont {Kurth}, \citenamefont {Rubio},\ and\
  \citenamefont {{E.K.U.~Gross}}}]{StefanucciKurthRubioGross:06}%
  \BibitemOpen
  \bibfield  {author} {\bibinfo {author} {\bibfnamefont {G.}~\bibnamefont
  {Stefanucci}}, \bibinfo {author} {\bibfnamefont {S.}~\bibnamefont {Kurth}},
  \bibinfo {author} {\bibfnamefont {A.}~\bibnamefont {Rubio}}, \ and\ \bibinfo
  {author} {\bibnamefont {{E.K.U.~Gross}}},\ }in\ \href@noop {} {\emph
  {\bibinfo {booktitle} {Molecular and Nano Electronics: Analysis, Design, and
  Simulation, 17}}},\ \bibinfo {editor} {edited by\ \bibinfo {editor}
  {\bibfnamefont {J.}~\bibnamefont {Seminario}}}\ (\bibinfo  {publisher}
  {Elsevier},\ \bibinfo {address} {Amsterdam},\ \bibinfo {year}
  {2006})\BibitemShut {NoStop}%
\bibitem [{\citenamefont {Vignale}\ and\ \citenamefont {{Di
  Ventra}}(2009)}]{VignaleDiVentra:09}%
  \BibitemOpen
  \bibfield  {author} {\bibinfo {author} {\bibfnamefont {G.}~\bibnamefont
  {Vignale}}\ and\ \bibinfo {author} {\bibfnamefont {M.}~\bibnamefont {{Di
  Ventra}}},\ }\href {\doibase 10.1103/PhysRevB.79.014201} {\bibfield
  {journal} {\bibinfo  {journal} {Phys. Rev. B}\ }\textbf {\bibinfo {volume}
  {79}},\ \bibinfo {pages} {014201} (\bibinfo {year} {2009})}\BibitemShut
  {NoStop}%
\bibitem [{\citenamefont {Stefanucci}\ and\ \citenamefont
  {Kurth}(2015)}]{Stefanucci:NL:2015}%
  \BibitemOpen
  \bibfield  {author} {\bibinfo {author} {\bibfnamefont {G.}~\bibnamefont
  {Stefanucci}}\ and\ \bibinfo {author} {\bibfnamefont {S.}~\bibnamefont
  {Kurth}},\ }\href {\doibase 10.1021/acs.nanolett.5b03294} {\bibfield
  {journal} {\bibinfo  {journal} {Nano Lett.}\ }\textbf {\bibinfo {volume}
  {15}},\ \bibinfo {pages} {8020} (\bibinfo {year} {2015})}\BibitemShut
  {NoStop}%
\bibitem [{\citenamefont {Yang}\ \emph {et~al.}(2016)\citenamefont {Yang},
  \citenamefont {Perfetto}, \citenamefont {Kurth}, \citenamefont {Stefanucci},\
  and\ \citenamefont {D'Agosta}}]{Yang:PRB:2016}%
  \BibitemOpen
  \bibfield  {author} {\bibinfo {author} {\bibfnamefont {K.}~\bibnamefont
  {Yang}}, \bibinfo {author} {\bibfnamefont {E.}~\bibnamefont {Perfetto}},
  \bibinfo {author} {\bibfnamefont {S.}~\bibnamefont {Kurth}}, \bibinfo
  {author} {\bibfnamefont {G.}~\bibnamefont {Stefanucci}}, \ and\ \bibinfo
  {author} {\bibfnamefont {R.}~\bibnamefont {D'Agosta}},\ }\href@noop {}
  {\bibfield  {journal} {\bibinfo  {journal} {Phys. Rev. B}\ }\textbf {\bibinfo
  {volume} {94}},\ \bibinfo {pages} {081410} (\bibinfo {year}
  {2016})}\BibitemShut {NoStop}%
\bibitem [{\citenamefont {Kurth}\ and\ \citenamefont
  {Stefanucci}(2016)}]{Kurth:PRB:2016}%
  \BibitemOpen
  \bibfield  {author} {\bibinfo {author} {\bibfnamefont {S.}~\bibnamefont
  {Kurth}}\ and\ \bibinfo {author} {\bibfnamefont {G.}~\bibnamefont
  {Stefanucci}},\ }\href {\doibase 10.1103/PhysRevB.94.241103} {\bibfield
  {journal} {\bibinfo  {journal} {Phys. Rev. B}\ }\textbf {\bibinfo {volume}
  {94}},\ \bibinfo {pages} {241103} (\bibinfo {year} {2016})}\BibitemShut
  {NoStop}%
\bibitem [{\citenamefont {Jacob}\ and\ \citenamefont
  {Kurth}(2018)}]{JacobKurth:18}%
  \BibitemOpen
  \bibfield  {author} {\bibinfo {author} {\bibfnamefont {D.}~\bibnamefont
  {Jacob}}\ and\ \bibinfo {author} {\bibfnamefont {S.}~\bibnamefont {Kurth}},\
  }\href@noop {} {\bibfield  {journal} {\bibinfo  {journal} {Nano Lett.}\
  }\textbf {\bibinfo {volume} {18}},\ \bibinfo {pages} {2086} (\bibinfo {year}
  {2018})}\BibitemShut {NoStop}%
\bibitem [{Note1()}]{Note1}%
  \BibitemOpen
  \bibinfo {note} {Needless to say, we assume $T>\Delta T/2$ such that
  $T_{L},T_{R}>0$.}\BibitemShut {Stop}%
\bibitem [{\citenamefont {Kurth}\ \emph {et~al.}(2019)\citenamefont {Kurth},
  \citenamefont {Jacob}, \citenamefont {Sobrino},\ and\ \citenamefont
  {Stefanucci}}]{KurthJacobSobrinoStefanucci:19}%
  \BibitemOpen
  \bibfield  {author} {\bibinfo {author} {\bibfnamefont {S.}~\bibnamefont
  {Kurth}}, \bibinfo {author} {\bibfnamefont {D.}~\bibnamefont {Jacob}},
  \bibinfo {author} {\bibfnamefont {N.}~\bibnamefont {Sobrino}}, \ and\
  \bibinfo {author} {\bibfnamefont {G.}~\bibnamefont {Stefanucci}},\ }\href
  {\doibase 10.1103/PhysRevB.100.085114} {\bibfield  {journal} {\bibinfo
  {journal} {Phys. Rev. B}\ }\textbf {\bibinfo {volume} {100}},\ \bibinfo
  {pages} {085114} (\bibinfo {year} {2019})}\BibitemShut {NoStop}%
\bibitem [{\citenamefont {Dittmann}\ \emph {et~al.}(2018)\citenamefont
  {Dittmann}, \citenamefont {Splettstoesser},\ and\ \citenamefont
  {Helbig}}]{dittmann:PRL:2018}%
  \BibitemOpen
  \bibfield  {author} {\bibinfo {author} {\bibfnamefont {N.}~\bibnamefont
  {Dittmann}}, \bibinfo {author} {\bibfnamefont {J.}~\bibnamefont
  {Splettstoesser}}, \ and\ \bibinfo {author} {\bibfnamefont {N.}~\bibnamefont
  {Helbig}},\ }\href@noop {} {\bibfield  {journal} {\bibinfo  {journal} {Phys.
  Rev. Lett.}\ }\textbf {\bibinfo {volume} {120}},\ \bibinfo {pages} {157701}
  (\bibinfo {year} {2018})}\BibitemShut {NoStop}%
\bibitem [{\citenamefont {Dittmann}\ \emph {et~al.}(2019)\citenamefont
  {Dittmann}, \citenamefont {Helbig},\ and\ \citenamefont
  {Kennes}}]{DittmannHelbigKennes:19}%
  \BibitemOpen
  \bibfield  {author} {\bibinfo {author} {\bibfnamefont {N.}~\bibnamefont
  {Dittmann}}, \bibinfo {author} {\bibfnamefont {N.}~\bibnamefont {Helbig}}, \
  and\ \bibinfo {author} {\bibfnamefont {D.~M.}\ \bibnamefont {Kennes}},\
  }\href {\doibase 10.1103/PhysRevB.99.075417} {\bibfield  {journal} {\bibinfo
  {journal} {Phys. Rev. B}\ }\textbf {\bibinfo {volume} {99}},\ \bibinfo
  {pages} {075417} (\bibinfo {year} {2019})}\BibitemShut {NoStop}%
\bibitem [{\citenamefont {Kurth}\ and\ \citenamefont
  {Stefanucci}(2017)}]{Kurth:JPCM:2017}%
  \BibitemOpen
  \bibfield  {author} {\bibinfo {author} {\bibfnamefont {S.}~\bibnamefont
  {Kurth}}\ and\ \bibinfo {author} {\bibfnamefont {G.}~\bibnamefont
  {Stefanucci}},\ }\href {\doibase 10.1088/1361-648X/aa7e36} {\bibfield
  {journal} {\bibinfo  {journal} {J. Phys. Condens. Mat.}\ }\textbf {\bibinfo
  {volume} {29}},\ \bibinfo {pages} {413002} (\bibinfo {year}
  {2017})}\BibitemShut {NoStop}%
\bibitem [{\citenamefont {Kurth}\ and\ \citenamefont
  {Stefanucci}(2018)}]{KurthStefanucci:18}%
  \BibitemOpen
  \bibfield  {author} {\bibinfo {author} {\bibfnamefont {S.}~\bibnamefont
  {Kurth}}\ and\ \bibinfo {author} {\bibfnamefont {G.}~\bibnamefont
  {Stefanucci}},\ }\href@noop {} {\bibfield  {journal} {\bibinfo  {journal}
  {Eur. J. Phys. B}\ }\textbf {\bibinfo {volume} {91}},\ \bibinfo {pages} {118}
  (\bibinfo {year} {2018})}\BibitemShut {NoStop}%
\bibitem [{Note2()}]{Note2}%
  \BibitemOpen
  \bibinfo {note} {Notice that by comparing (18) and (9) one arrives at
  $S=S_s-S_{xc}$. This reflects the standard definition
  $V_{xc}=V_s-V$.}\BibitemShut {Stop}%
\bibitem [{\citenamefont {Costi}\ and\ \citenamefont
  {Zlati{\'c}}(2010)}]{costi:PRB:2010}%
  \BibitemOpen
  \bibfield  {author} {\bibinfo {author} {\bibfnamefont {T.}~\bibnamefont
  {Costi}}\ and\ \bibinfo {author} {\bibfnamefont {V.}~\bibnamefont
  {Zlati{\'c}}},\ }\href@noop {} {\bibfield  {journal} {\bibinfo  {journal}
  {Phys. Rev. B}\ }\textbf {\bibinfo {volume} {81}},\ \bibinfo {pages} {235127}
  (\bibinfo {year} {2010})}\BibitemShut {NoStop}%
\bibitem [{\citenamefont {Stefanucci}\ and\ \citenamefont
  {Kurth}(2011)}]{StefanucciKurth:11}%
  \BibitemOpen
  \bibfield  {author} {\bibinfo {author} {\bibfnamefont {G.}~\bibnamefont
  {Stefanucci}}\ and\ \bibinfo {author} {\bibfnamefont {S.}~\bibnamefont
  {Kurth}},\ }\href@noop {} {\bibfield  {journal} {\bibinfo  {journal} {Phys.
  Rev. Lett.}\ }\textbf {\bibinfo {volume} {107}},\ \bibinfo {pages} {216401}
  (\bibinfo {year} {2011})}\BibitemShut {NoStop}%
\bibitem [{\citenamefont {Bergfield}\ \emph {et~al.}(2012)\citenamefont
  {Bergfield}, \citenamefont {Liu}, \citenamefont {Burke},\ and\ \citenamefont
  {Stafford}}]{Bergfield:PRL:2012}%
  \BibitemOpen
  \bibfield  {author} {\bibinfo {author} {\bibfnamefont {J.~P.}\ \bibnamefont
  {Bergfield}}, \bibinfo {author} {\bibfnamefont {Z.-F.}\ \bibnamefont {Liu}},
  \bibinfo {author} {\bibfnamefont {K.}~\bibnamefont {Burke}}, \ and\ \bibinfo
  {author} {\bibfnamefont {C.~A.}\ \bibnamefont {Stafford}},\ }\href {\doibase
  10.1103/PhysRevLett.108.066801} {\bibfield  {journal} {\bibinfo  {journal}
  {Phys. Rev. Lett.}\ }\textbf {\bibinfo {volume} {108}},\ \bibinfo {pages}
  {066801} (\bibinfo {year} {2012})}\BibitemShut {NoStop}%
\bibitem [{\citenamefont {Tr\"oster}\ \emph {et~al.}(2012)\citenamefont
  {Tr\"oster}, \citenamefont {Schmitteckert},\ and\ \citenamefont
  {Evers}}]{TroesterSchmitteckertEvers:12}%
  \BibitemOpen
  \bibfield  {author} {\bibinfo {author} {\bibfnamefont {P.}~\bibnamefont
  {Tr\"oster}}, \bibinfo {author} {\bibfnamefont {P.}~\bibnamefont
  {Schmitteckert}}, \ and\ \bibinfo {author} {\bibfnamefont {F.}~\bibnamefont
  {Evers}},\ }\href {\doibase 10.1103/PhysRevB.85.115409} {\bibfield  {journal}
  {\bibinfo  {journal} {Phys. Rev. B}\ }\textbf {\bibinfo {volume} {85}},\
  \bibinfo {pages} {115409} (\bibinfo {year} {2012})}\BibitemShut {NoStop}%
\bibitem [{\citenamefont {Jakobs}\ \emph {et~al.}(2010)\citenamefont {Jakobs},
  \citenamefont {Pletyukhov},\ and\ \citenamefont
  {Schoeller}}]{Jakobs:PRB:2010}%
  \BibitemOpen
  \bibfield  {author} {\bibinfo {author} {\bibfnamefont {S.~G.}\ \bibnamefont
  {Jakobs}}, \bibinfo {author} {\bibfnamefont {M.}~\bibnamefont {Pletyukhov}},
  \ and\ \bibinfo {author} {\bibfnamefont {H.}~\bibnamefont {Schoeller}},\
  }\href {\doibase 10.1103/PhysRevB.81.195109} {\bibfield  {journal} {\bibinfo
  {journal} {Phys. Rev. B}\ }\textbf {\bibinfo {volume} {81}},\ \bibinfo
  {pages} {195109} (\bibinfo {year} {2010})}\BibitemShut {NoStop}%
\bibitem [{\citenamefont {Izumida}\ \emph {et~al.}(2001)\citenamefont
  {Izumida}, \citenamefont {Sakai},\ and\ \citenamefont
  {Suzuki}}]{Izumida:PSJ:2001}%
  \BibitemOpen
  \bibfield  {author} {\bibinfo {author} {\bibfnamefont {W.}~\bibnamefont
  {Izumida}}, \bibinfo {author} {\bibfnamefont {O.}~\bibnamefont {Sakai}}, \
  and\ \bibinfo {author} {\bibfnamefont {S.}~\bibnamefont {Suzuki}},\
  }\href@noop {} {\bibfield  {journal} {\bibinfo  {journal} {J. Phys. Soc.
  Jpn}\ }\textbf {\bibinfo {volume} {70}},\ \bibinfo {pages} {1045} (\bibinfo
  {year} {2001})}\BibitemShut {NoStop}%
\end{thebibliography}
\end{document}